\documentclass[prd,reprint,amsmath,aps,superscriptaddress,nofootinbib]{revtex4-1}
\usepackage[colorlinks,linkcolor={blue},citecolor={blue}]{hyperref}
 \usepackage{graphicx}
\usepackage{xcolor}
\usepackage{bm}
\usepackage{acronym}
\usepackage{graphicx}

\begin{document}

\title{Measuring gravitational-wave higher-order modes}

\author{Cameron Mills}
\affiliation{Gravity Exploration Institute, School of Physics and Astronomy, Cardiff University, The Parade, Cardiff,
CF24 3AA, UK}
\author{Stephen Fairhurst}
\affiliation{Gravity Exploration Institute, School of Physics and Astronomy, Cardiff University, The Parade, Cardiff,
CF24 3AA, UK}

\begin{abstract}
We investigate the observability of higher harmonics in gravitational wave signals emitted during the coalescence of binary black holes.
We decompose each mode into an overall amplitude, dependent upon the masses and spins of the system, and an 
orientation-dependent term, dependent upon the inclination and polarization of the source.  Using this decomposition, we investigate
the significance of higher modes over the parameter space and show that the $\ell = 3$, $m = 3$ mode is most significant across much of the sensitive 
band of ground-based interferometric detectors, with the $\ell = 4$, $m = 4$ having a significant contribution at high masses.  We introduce the higher mode signal-to-noise ratio (SNR), and show that a simple threshold on this SNR can be used as a criterion for observation of higher harmonics.  Finally, we investigate observability in a population of binaries and observe that higher harmonics will only be observable in a few percent of binaries, typically those with unequal masses and viewed close to edge-on.  
\end{abstract}

\maketitle


\acrodef{GW}{gravitational-wave}
\acrodef{LIGO}{Laser Interferometer Gravitational-wave Observatory}
\acrodef{BBH}{binary black hole}
\acrodef{O1}{first observing run}
\acrodef{O2}{second observing run}
\acrodef{O3}{third observing run}
\acrodef{BH}{black hole}
\acrodef{SNR}{signal-to-noise ratio}
\acrodef{BNS}{binary neutron star}
\acrodef{PSD}{power spectral density}
\acrodef{GR}{general relativity}
\acrodef{PPD}{posterior predictive distribution}
\acrodef{FAR}{false-alarm rate}
\acrodef{PN}{post-Newtonian}

\newcommand\red[1]{{\color[rgb]{0.75,0.0,0.0} #1}}
\newcommand\green[1]{{\color[rgb]{0.0,0.60,0.08} #1}}
\newcommand\blue[1]{{\color[rgb]{0,0.20,0.65} #1}}
\newcommand{\roberto}[1]{\textcolor{magenta}{(Roberto:#1)}}
\newcommand\numberthis{\addtocounter{equation}{1}\tag{\theequation}}
\date{\today}
\def\LL#1{\blue{(Lionel: #1)}}
\def\fig#1{Fig.~\ref{#1}}
\newcommand{\figs}[2]{Figures~(\ref{#1}-\ref{#2})}
\def\eqn#1{Eq.~(\ref{#1})}
\def\sec#1{Sec.~\ref{#1}}
\newcommand{\eqns}[2]{Eqs.~(\ref{#1}-\ref{#2})}
\def\hp{h_+}
\def\hc{h_\times}
\def\ylm{\prescript{}{-2}{Y_{\ell m}}(\theta,\phi)}
\def\lm{\ell m}
\def\hlm{h_{\lm}}

\let\apjl=\apj
\def\araa{Annu. Rev. Astron. Astrophys.}

\newcommand{\camcomment}[1]{\textcolor{green}{[Cam: #1]}}
\newcommand{\steve}[1]{\textcolor{magenta}{[Steve: #1]}}
\newcommand{\thesis}[1]{\textcolor{blue}{[Keep for thesis: #1]}}
\newcommand{\checkme}[1]{\textcolor{red}{#1}}


\section{Introduction}

Gravitational waves emitted during the coalescence of black hole and/or neutron star binaries are emitted predominantly 
at twice the orbital frequency, during the inspiral phase of the coalescence \citep{Blanchet:2013haa}.  However, it is 
also well-known that the gravitational wave signal cannot be completely characterized by a single harmonic but, rather, 
is better decomposed as a sum of spin-weighed spherical \citep{Thorne:1980ru} (or spheroidal \citep{Leaver:1985ax}) 
harmonics.  The dominant harmonic is the $\ell = 2$, $m = \pm 2$ harmonic, but there is also power in higher harmonics, 
most notably the 21, 32, 33 and 44 harmonics
\footnote{
	When we refer to a mode by the label $\ell$m we mean $\ell$, $\pm m$.
}
 \citep{Kidder:2007rt, Mishra:2016whh}.  The importance of these additional 
harmonics increases as the mass ratio between the two black holes increases and also increases during the late inspiral 
and merger of the objects.  Recent semi-analytical and numerical relativity models have provided expressions for an 
increasing number of the higher harmonics accurate across the inspiral, merger and ringdown regimes 
\citep{Mehta:2017jpq, London:2017bcn, Khan:2018fmp, Khan:2019kot, Mehta:2019wxm, Cotesta:2018fcv, Varma:2018mmi, 
Varma:2019csw, Rifat:2019ltp, Nagar:2020pcj, Garcia-Quiros:2020qpx, Cotesta:2020qhw, Ossokine:2020, Pratten:2020ceb}.

Clear evidence of higher gravitational-wave harmonics has been observed in two recent observations, GW190412 
\citep{LIGOScientific:2020stg} and GW190814 \citep{Abbott:2020khf}, as well as weaker evidence in the high-mass system 
GW170729 \citep{Chatziioannou:2019dsz}.  These observations provide further evidence that Einstein's general relativity 
is an accurate description of gravity, including in the strong-field, highly dynamic regime of the merger of two black 
holes \citep{Kidder:2007rt, VanDenBroeck:2006qu}.  By incorporating knowledge of the higher harmonics into a search for 
gravitational waves, the sensitivity of gravitational wave searches can be improved, leading to an increase in the rate 
of observed systems \citep{Harry:2017weg}; furthermore these observations would typically be from less densely populated 
regions of the parameter space \citep{O2RatesPops}, for example high mass binaries and those with unequal mass components.  
Finally, the observation of higher harmonics enables more accurate measurement of the properties of system 
\citep{Kalaghatgi:2019log, LIGOScientific:2020stg, Abbott:2020khf}.  For example, the measurement of multiple harmonics 
can be used to break well-known degeneracies between the measured distance and orientation of the system \citep{Usman:2018imj}, 
or the mass ratio and spins of the black holes \citep{Ohme:2013nsa, Hannam:2013uu}.

While the gravitational waveform is comprised of an infinite number of harmonics, it is the unambiguous measurement of 
a \textit{second harmonic} (in addition to the 22-harmonic) which will lead to a step-change in our ability to measure 
the properties of the system; additional harmonics will then further refine the measurement accuracy.  In this paper, 
we perform an in-depth investigation of the importance of the higher harmonics across the parameter space and identify 
regions of the parameter space where particular harmonics are most likely to make a significant contribution. The 
amplitude of each harmonic depends both upon the intrinsic parameters of the system (its masses and spins, both 
magnitudes and orientations) as well as the extrinsic parameters (the orientation of the binary and the detector 
network's sensitivity to the two polarizations of gravitational waves).  For simplicity, we decompose the harmonics 
into an overall amplitude factor, dependent only upon the intrinsic parameters, and an orientation dependent term.  
We then investigate the significance of each harmonic across the parameter space.

Next, we turn to the question of \textit{when} additional harmonics have been unambiguously observed. From a model selection 
perspective, this can be addressed by considering the evidence in favour of a waveform containing higher harmonics against one
without.  Here, we introduce the higher-harmonic signal to noise ratio, and argue that it can be used as an alternative method
of establishing the observability of higher harmonics.%
\footnote{A similar prescription has recently been introduced for precessing systems \citep{Fairhurst:2019srr,Fairhurst:2019vut}.}
It is straightforward to calculate the SNR contained in each of the higher waveform harmonics, and compare to the expectation 
due to noise-only in the higher harmonics.  This approach has been used to verify the observation of higher harmonics from 
the binary mergers observed as GW190412 and GW190814 \citep{LIGOScientific:2020stg, Abbott:2020khf}.

The structure of the paper is as follows.  In section \ref{sec:waveform}, we provide a brief review of the gravitational 
waveform, incorporating the higher harmonics, and use this to fix the notation for the remainder of the paper; in section 
\ref{sec:significance} we explore the significance of the higher harmonics over the parameter space, both intrinsic 
(masses and spins) and extrinsic (binary orientation); in section \ref{sec:observing} we investigate the observability 
of higher harmonics and introduce a simple criterion for detection; finally in section \ref{sec:population} we investigate 
observability for a population of events.

%
%
%

\section{The Gravitational Waveform}
\label{sec:waveform}
The measured gravitational wave strain $h$ can be written as 
\begin{equation}
h = F_+ h_+ + F_{\times} h_{\times}, 
\label{eq:detector_response}
\end{equation}
where the antenna factors $F_{+}$ and $F_{\times}$ depend upon the sky location (right-ascension and declination) of the source, 
as well as the polarization of the source.  It is often convenient to explicitly extract the polarization angle $\psi$ and then consider the
detector response to be a \textit{known} quantity dependent upon only the details of the detector and the direction to
the source.  Thus, we write the detector response as,
\begin{eqnarray}
	F_{+} &=& w_{+} \cos 2\psi + w_{\times} \sin 2\psi, \nonumber \\
	F_{\times} &=& - w_{+} \sin 2\psi + w_{\times} \cos 2\psi \label{eq:f_plus_cross},
\end{eqnarray}
where $w_{+}$ and $w_{\times}$ are the detector response functions in a fixed frame --- for a single detector it
is natural to choose $w_{\times} = 0$ and for a network to work in the dominant polarization, in which for each sky 
point the polarization angle $\chi$ is chosen to maximize the network sensitivity to $w_{+}$ \citep{Klimenko:2005xv,Harry:2010fr}.  
The relative amplitude of $w_{\times}$ to $w_{+}$ describes the sensitivity of the network to the second gravitational wave polarization.  
The unknown polarization of the source relative to this preferred frame is denoted $\psi$.

%

The radiation-frame gravitational wave polarizations $h_+$ and $h_{\times}$ can be decomposed into modes using spin-weighted spherical harmonics of spin weight $-2$, ${}_{-2}Y_{lm}$, which are functions of the inclination angle $\iota$ and coalescence phase $\phi_c$ (see Appendix \ref{sec:Ylm} for a more detailed discussion of the decomposition).
%
For a binary merger which does not exhibit precession, the waveform can be expressed in the frequency domain, using the
stationary-phase approximation, as 
\begin{align}\label{eq:Alm_polarizations}
h_+ &=  \frac{d_{o}}{d_{L}} \sum_{l \geq 2} \sum_{m=0}^{l} A_{+}^{lm}(\iota) e^{i m \phi_{c}} \tilde{h}_{lm}(f) \\
h_{\times} &= \frac{d_{o}}{d_{L}}  \sum_{l \geq 2}\sum_{m=0}^{l} A_{\times}^{lm}(\iota) i e^{i m \phi_{c}} \tilde{h}_{lm}(f) \nonumber
\end{align}
where $d_{L}$ is the luminosity distance, $d_{o}$ is a fiducial distance used to normalize the waveforms $\tilde{h}_{\ell m}$.  The 
amplitudes $A^{\ell m}$ are functions only of the inclination angle and are given below for the most significant harmonics:
%
%
\begin{align}\label{eq:iota_mode_dependence}
A^{22}_+ &=  \tfrac{1}{2}(1+\cos^2 \iota) \\
A^{22}_\times &= \cos \iota \nonumber \\
A^{21}_+ &= \sin \iota \nonumber\\
A^{21}_\times &=  \sin \iota \cos \iota \nonumber\\
A^{33}_+ &= \sin \iota (1+\cos^2 \iota) \nonumber\\ 
A^{33}_\times &= 2 \sin \iota \cos \iota \nonumber \\
A^{32}_+ &= 1 - 2 \cos^2 \iota \nonumber \\
A^{32}_\times &= \tfrac{1}{2} (\cos \iota - 3 \cos^3 \iota) \nonumber \\
A^{44}_+ &= \sin^2 \iota (1+\cos^2 \iota) \nonumber \\
A^{44}_\times &= 2 \sin^2 \iota \cos \iota \nonumber 
\end{align}
There is a freedom in choice of overall normalization for these amplitudes, which corresponds to an overall rescaling of the waveform defining each mode, $\tilde{h}_{\ell m}$.  For the 22 mode, it is customary to choose a normalization such that $A^{22}_{+} = A^{22}_{\times} = 1$ for a face-on system, and we use that normalization here.  Since many of the higher harmonics vanish for face-on systems, we instead choose a normalization for the higher-mode amplitudes, $A^{lm}_{+, \times}$ in Eq.~(\ref{eq:iota_mode_dependence}), by requiring that for theplus polarization $A^{lm}_{+} = 1$ at $\iota = \tfrac{\pi}{2}$, i.e. when the system is edge on.  

\begin{figure*}[t]
\includegraphics[width=0.49\linewidth]{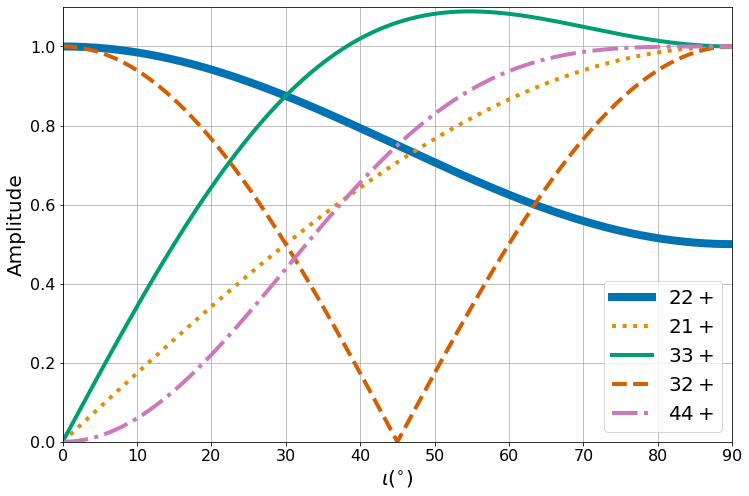}
\includegraphics[width=0.49\linewidth]{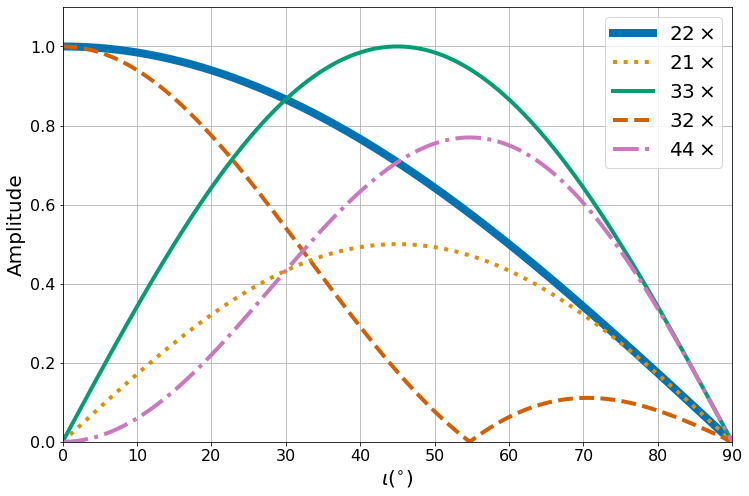}
\caption{The amplitude of the 22, 21, 33, 32 and 44 harmonics as a function of the inclination $\iota$ of the binary.  The 22 mode is normalized to unity at $\iota = 0^{\circ}$ while other modes are normalized to unit amplitude in the $+$ polarization at $\iota = 90^{\circ}$. \textit{Left:} + polarization \textit{Right:} $\times$ polarization. 
}
\label{fig:iota_dependence}
\end{figure*}

\fig{fig:iota_dependence} shows the dependence of the modes on inclination. The plus polarization of the 22 mode peaks at face-on, while the 21 and 44 modes peak at edge-on. The 32 mode amplitude is maximum at both face-on and edge-on orientations while the 33 mode peaks at $\sin \iota =\sqrt{\tfrac{2}{3}}$.  The different dependence of the modes on the binary orientation can lead to the improved measurement of the inclination, when more than one mode is observed \citep{LIGOScientific:2020stg, Kalaghatgi:2019log}, breaking the well-known degeneracy between distance and inclination angle that arises when observing only the dominant harmonic \citep{Usman:2018imj}.

During inspiral the frequency evolution of a multipole $\omega_{lm}$, is related to the orbital frequency $\omega_{orb}$ as $\omega_{lm} \sim m \omega_{orb}$ \citep{Kidder:2007rt}. While during the ringdown the frequency approximately evolves as $\omega_{lm} \sim l \omega_{orb}$ \citep{Leaver:1985ax, Berti:2005ys}.
Thus it is possible to scale the frequencies of the 22 mode in quite a simple manner to obtain an approximate phase evolution of the $l=m$ harmonics, for example the phase evolution of the 33 mode is approximately a factor of 1.5 times $\omega_{22}$ \cite{Roy:2019phx}.


\section{The Significance of Higher Harmonics}
\label{sec:significance}
The gravitational wave signal from every binary merger will be comprised of the sum of harmonics.  However, for the majority of signals observed close to threshold, only the dominant 22 harmonic will be observable above the noise background.  In this section, we investigate the observability of the different modes, and how this varies across the mass and spin parameter space.
 For concreteness, we restrict attention to a single detector with a sensitivity comparable to that achieved by the LIGO observatories during their third observing run \citep{PSDs}.

The key metric for waveform observability is the optimal \ac{SNR} defined as 
\begin{equation}
\label{eq:SNR}
\hat{\rho}=\sqrt{(h|h)} \, ,
\end{equation}
where we have introduced the inner product weighted by noise characterized by a power spectrum $S(f)$
\begin{equation}
\label{eq:m_filter}
\left( a| b\right) := 4 \, \mathrm{Re} \int_{0}^{f_{\mathrm{max}}} \frac{\tilde{a}(f)\tilde{b}(f)^{\star}}{S(f)} \; \text{d}f \, . 
\end{equation}

Consider the situation where the 22 mode has been observed, and we are interested in obtaining an estimate of the expected SNR in the other harmonics.  As is clear from \eqn{eq:Alm_polarizations}, the SNR in the higher harmonics will depend upon the detector sensitivity to the higher harmonic waveform, $\tilde{h}_{\ell m}$, as well as the amplitude factor $A^{\ell m}_{+,\times}$.  

Let us examine the single-detector case in detail.  For simplicity, we choose a detector sensitive only to the $+$ polarization (in the 
preferred frame), so that $w_{\times} = 0$, and we take $w_{+} = 1$.  Furthermore, we simplify the calculation to consider only two 
modes, the dominant $(2, 2)$ harmonic and one other generic $\ell m$ harmonic.  The amplitude of each multipole depends on both the 
intrinsic properties of the system and the orientation relative to the network of detectors. 
The waveform observed at the detector is
\begin{align}\label{eq:two_harmonics}
h 
&= \cos 2\psi (h_{+}^{22} + h_{+}^{\ell m})  - \sin 2 \psi (h_{\times}^{22} + h_{\times}^{\ell m}) \, , 
\end{align}
where we have implicitly defined $h_{+, \times}^{\ell m}$ as the $\ell m$ component of $h_+$ and $h_{\times}$ in \eqn{eq:Alm_polarizations}.

From this, we can calculate the optimal SNR as
\begin{align*}
\hat{\rho}^{2} = & \cos^{2} 2\psi \left[ |h_+^{22}|^2 + |h_+^{\ell m}|^2 + 2 (h_+^{\ell m}|h_+^{22}) \right] \\
	+& \sin^{2} 2\psi \left[ |h_\times^{22}|^2 + |h_\times^{\ell m}|^2 + 2 (h_\times^{\ell m}|h_\times^{22}) \right] 
	\numberthis \label{eq:h}
\end{align*}
where the cross terms ($\sin 2\psi \cos 2\psi$) cancel since $(a|ib)=-(ia|b)$.


The cross terms between modes, $(h_{+, \times}^{\ell m}|h_{+, \times}^{22})$ can be both positive or negative, causing constructive or destructive interference 
between the harmonics. As discussed previously, the frequency during inspiral scales with $m$ while the ringdown frequency 
scales approximately with $\ell$.  Consequently, there is typically little overlap between the 22 mode and modes for which 
both $\ell \ne 2$ and $m \ne 2$.  Since the most signficant subdominant multipoles are usually the 33 and 44 harmonics we 
neglect these cross terms for now, but will revisit their significance later. Doing this allows us to write
\begin{align*}
   \hat{\rho}^{2} = & \cos^{2} 2\psi  |h_+^{22}|^2 \left[ 1 + \frac{|h_+^{\ell m}|^2}{|h_+^{22}|^2} \right] \\
    +& \sin^{2} 2\psi  |h_\times^{22}|^2 \left[ 1 + \frac{|h_\times^{\ell m}|^2}{|h_\times^{22}|^2 } \right] \, .
    \numberthis \label{eq:rho2}
\end{align*}
This enables us to capture the relative importance of the $\ell m$ harmonic in two terms, which are the ratio of the amplitude of the $\ell m$ harmonic ($+$ and $\times$ components) to the dominant, 22 harmonic.  This ratio depends upon both the intrinsic amplitude of the two modes, as well as the orientation-dependent amplitude of the mode.  Therefore, we can define
\begin{equation}
\frac{|h_{+, \times}^{\ell m}|}{|h_{+,\times}^{22}|} = \alpha_{\ell m} R^{+, \times}_{\ell m}
\end{equation}
where
\begin{equation}\label{eq:alpha_lm}
    \alpha_{\ell m} = 
    \sqrt{\frac{(\tilde{h}_{\ell m }| \tilde{h}_{\ell m })}{(\tilde{h}_{22}|\tilde{h}_{22})} \, }.
\end{equation}
encodes the relative amplitude of the modes and
\begin{equation}\label{eq:Rpluscross}
R^{+, \times}_{\ell m} = \frac{|A^{\ell m}_{+, \times}|}{|A^{22}_{+, \times}|}
\end{equation}
encodes the relative size of the orientation factors.
%
%

We can thus write the \ac{SNR}, neglecting the overlap between modes, as
\begin{equation}
    \rho^2 = \rho_{22}^2 + \rho_{\ell m}^2 \, ,
\end{equation}
where 
\begin{equation}\label{eq:rho_lm}
\rho_{\ell m}  = \rho_{22} \alpha_{\ell m} R_{\ell m} \, .
\end{equation}
In general, the relative mode amplitudes $R_{\ell m}$ will depend upon both the inclination and polarization angles.  However,
for the $\ell = m$ modes, things simplify as the orientation amplitudes for $+, \times$ are the same.  In this case, there is no 
dependence upon the polarization angle and
\begin{align*}
\label{eq:Rll}
&R_{33}(\iota) = 2 \sin \iota \\
&R_{44}(\iota) = 2 \sin^2 \iota \, .
\numberthis
\end{align*}
%

In the remainder of this section, we explore the dependence of the relative amplitudes $\alpha_{\ell m}$ over the mass and spin parameter space and the expected distribution of $R_{\ell m}$ for a population of sources.

\subsection{Dependence upon intrinsic parameters}
\begin{figure*}[t] 
\includegraphics[width=0.99\textwidth]{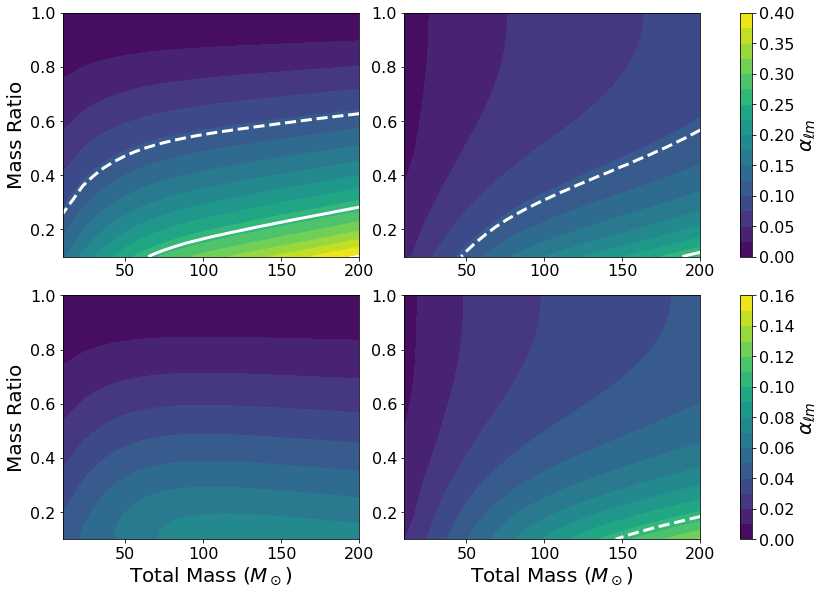}
\caption{Ratio of the intrinsic amplitude, $\alpha_{lm}$, (defined in equation (\ref{eq:alpha_lm})) of signal harmonics to the 22 harmonic as a function of the total (detector frame) mass and mass ratio of the system, in a detector with sensitivity matching the Advanced LIGO detectors during O3 \citep{PSDs}.  \textit{Upper left:} the 33 harmonic; \textit{Upper right:} the 44 harmonic; \textit{lower left:} the 21 harmonic; \textit{lower right:} the 32 harmonic.  In all cases, the spins of the black holes are set to zero.  The solid white line corresponds to $\alpha_{\ell m} = 5.3/20$ and the dashed line to $\alpha_{\ell m} = 2.1/20$, which correspond, approximately, to the threshold for the higher modes being confidently/marginally observable for a signal with SNR=20 in the 22 mode. Note that the colorbar is normalized differently between the top and bottom row to improve the visibility of the weaker modes.
}
\label{fig:alpha_lm}
\end{figure*}

The two important intrinsic parameters determining the relative power in the higher modes are mass ratio and total mass, with spin effects entering at higher \ac{PN} order for most modes \citep{Mishra:2016whh}. The contribution of a higher mode relative to the dominant 22 mode generically increases with an increasing mass ratio.  The relative amplitudes of the modes is independent of the total mass of the system.  However the frequency content of each mode does depend upon the total mass and thus depending on the shape of the detector power spectral density certain higher modes might be preferentially observed.  In particular, the contribution of higher modes can become more significant at high masses, for which the merger frequency of the dominant harmonic lies below the optimal sensitivity of the detector.

In \fig{fig:alpha_lm} we show the relative amplitude, $\alpha_{lm}$, of the four multipoles that we are considering: the 33, 44, 21 and 32 harmonics.  The amplitudes have been calculated using the PhenomHM waveform \citep{London:2017bcn}, for a signal observed in a detector with LIGO O3 sensitivity \citep{PSDs}, as a function of the (observed) total mass and mass ratio of the system.  Over much of the parameter space, the 33 mode will be the most significant, with the significance of the 33 mode increasing with mass ratio.  For example, at a total mass of $50 M_{\odot}$, the 33 mode has 10\% of the amplitude of the leading mode at a mass ratio of 2:1 and 20\% at 5:1.  At high masses, and significant mass ratios, the relative sensitivity to the 33 mode is greater than one third of the 22.  Over much of the parameter space, the 44 mode will be the third most significant mode.  However, sensitivity to the 44 mode increases rapidly as the mass of the system increases so that for total mass above $\sim 75 M_{\odot}$ and mass ratio less than 2:1, the 44 will be more significant than the 33.  The intrinsic amplitudes of the 21 and 32 modes are always less than the 33 and 44.  As with the other harmonics, these do become more significant as the mass ratio decreases and also (for the 32 mode in particular) as the total mass increases.  The 21 mode is the only subdominant mode considered in this paper which has spin terms in the amplitude at 1 \ac{PN} order \cite{Mishra:2016whh}. For this reason, the 21 mode can become more significant for binaries with large anti-aligned spins, and the intrinsic amplitude roughly doubles for a binary with effective spin $\chi_{\mathrm{eff}} = -0.8$.  Two facts determine the behaviour of $\alpha_{21}$ with increasing total mass: first, $\alpha_{21}$ is largest at merger and, second, the 21 merger is at a lower frequency than the 22. Initially $\alpha_{21}$ will increase with total mass as the detector becomes more sensitive to the merger.  However, as the mass continues to increase, the 21 mode amplitude is no longer observable as it merges at too low a frequency, and so $\alpha_{21}$ decreases.

\begin{figure*}[t] 
\includegraphics[width=0.99\linewidth]{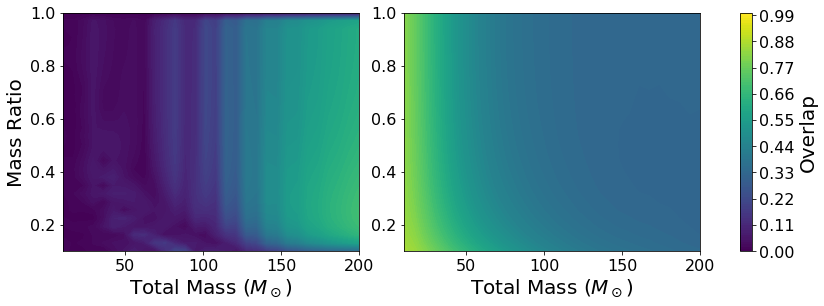} 
\caption{Absolute value of the noise-weighted inner product between multipoles, evaluated using the Advanced LIGO (O3) sensitivity, as a function of mass ratio and total mass for non-spinning black holes ($\chi_{\mathrm{eff}} = 0$).  \textit{Left:} the overlap between the 22 and 21 multipoles; \textit{Right:} the overlap between 22 and 32 multipoles.}
\label{fig:overlap_21_32}
\end{figure*}

For the power in these modes to be observable, it must be possible to distinguish the mode from the 22 harmonic.  Generally, it is only the contribution which is \textit{orthogonal} to the 22 harmonic which will be observable.  Any contribution from the higher harmonics which is proportional to the 22 harmonic will simply serve to change the power observed in the 22.  Consequently, we are interested in knowing whether the waveforms are orthogonal or, equivalently, what the overlap between the modes is.  Here, we define the normalized overlap maximized over $\phi_c$
\begin{equation}
O(\ell m, 22) = \frac{ \mathrm{Max}_{\phi_{c}} ( \tilde h_{\ell m} | \tilde{h}_{22})}{|\tilde{h}_{\ell m}| |\tilde{h}_{22}|} \, .
\end{equation}

The overlap between the 33 and 44 modes with the 22 harmonic is $<10\%$ across the parameter space explored, as expected due to the fact that the frequency evolution of these harmonics differs significantly from the 22.  However, the overlap of the 22 mode with the 21 and 32 modes can be significant.  These overlaps are shown in \fig{fig:overlap_21_32} as a function of total mass and mass ratio. As the 21 multipole has the same frequency as the 22 mode during ringdown, we expect a significant overlap at higher masses when the (merger and) ringdown occur within the sensitive band of the detector.  Similarly, for the 32 multipole, the frequency evolution during the inspiral matches closely with the 22 multipole and we therefore expect a significant overlap between the 22 and 32 modes, particularly for low masses.  Consequently, it can be difficult to observe these modes.  Interestingly, one of the most significant impact of the 32 mode can be to produce an incorrect estimate of the amplitude of the 22 mode, and consequently introduce an error in the measured distance, as power from the 32 mode will be mistakenly attributed to the 22 mode. \citep{VanDenBroeck:2006qu}

\subsection{Dependence upon extrinsic parameters}

\begin{figure*}[t]
\includegraphics[width=0.49\linewidth]{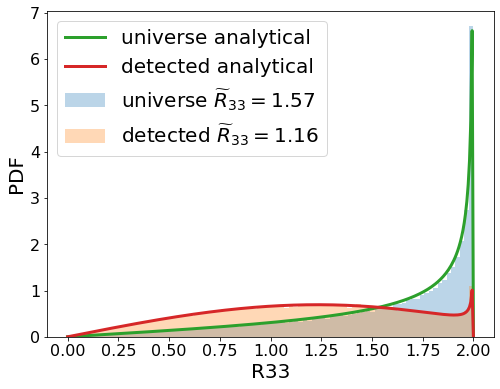}
\includegraphics[width=0.49\linewidth]{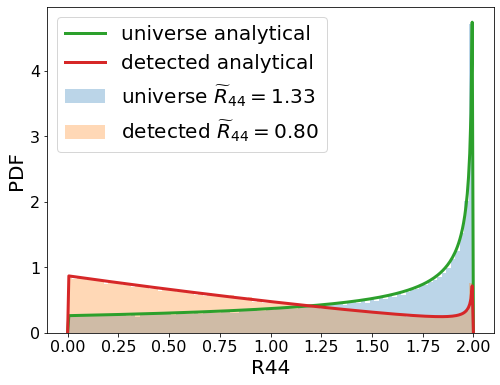}
\caption{Distribution of $R_{33}(\iota)$ and $R_{44}(\iota)$ for all binaries (Universe) as well as that subset that would be detected above a fixed SNR threshold for the 22 harmonic (Detected).  We show both the results from a Monte-Carlo simulation as well as the analytical prediction.}
\label{fig:R_dist}
\end{figure*}

The observed SNR in the higher harmonics depends upon the orientation of the binary, through the $R_{\ell m}$ factor (defined in \eqn{eq:Rpluscross}), in addition to the intrinsic amplitude of the modes discussed above.  We can make several immediate observations from \fig{fig:iota_dependence}.  The 33, 44 and 21 modes vanish for a signal observed face-on ($\iota = 0$), so the miminum value of $R_{\ell m}$ for these modes is zero; in contrast, there is no orientation for which both polarizations of the 32 mode vanishes.  Next, there is no orientation where the 22 mode vanishes, but the other modes do not --- the 22 mode only vanishes for the $\times$ polarization for an edge-on system, but all other modes we are considering also vanish there.  Thus, there is a finite, maximum value of $R_{\ell m}$ for all modes, and it's easy to see from \fig{fig:iota_dependence} that $R_{\ell m}^{\mathrm{max}} = 2$, which occurs for edge-on systems.

We now focus on the 33 and 44 modes: not only are these the most significant, as seen in \fig{fig:alpha_lm}, but also the expression for $R_{\ell m}$  simplifies as the relative amplitude of these modes is independent of the observed gravitational wave polarization.  We consider the distribution of $R_{\ell m}$ for a population of sources distributed uniformly in volume%
\footnote{Realistically, we do not expect sources to be uniformly distributed, due to both cosmological effects and 
a redshift dependent star formation and, hence, merger rate \citep{2014ARA&A..52..415M}.  
Nonetheless, this simple model provides a reasonably approximation to gain an understanding of the likely values of $R$.}
and with uniformly distributed orientation.  
In \fig{fig:R_dist} we plot the expected distribution of the geometrical factors $R_{33}(\iota)$ and $R_{44}(\iota)$. We show both the distribution based upon uniformly distributed sources, as well as the expected observed distribution --- 
obtained by placing a threshold upon the observed SNR in the dominant 22 harmonic \citep{Schutz:2011tw}.   For both modes, the distribution peaks at $R=2$, the value for an edge-on system.  However, since the emission in the 22 mode is weakest when the system is observed edge on, selection effects serve to significantly reduce the peak in the observed population.  For the 33 mode, the peak remains at $R_{33} = 2$, but the distribution is broad, with significant support over the full range from 0 to 2 and a mean value of $1.16$.  The mode of the observed $R_{44}$ distribution is zero, although again there is broad support over the range from 0 to 2 with a mean value of $0.8$.

For other modes, the expected distribution of $R_{\ell m}$ will depend upon the sensitivity of the detector network to the two polarizations of the gravitational wave --- the distribution for $R_{\ell m}$ will differ between a single detector, sensitive to only one polarization, and a network with good sensitivity to both polarizations.  
Nonetheless, the distribution for $R_{21}$ will share features with $R_{33}$ and $R_{44}$, namely it will take values between 0 (face on) 
and 2 (edge on), with a peak at $R_{21} = 2$ which is reduced by selection effects in the observed population.  The 32 mode is non-zero for a face-on signal, $R_{32} \lesssim 1$ for sources near to face-on, and so there is a significant fraction of sources with $R_{32} \approx 1$ as well as at the maximum of $R_{32} = 2$ which occurs for edge-on systems.

\section{Observing Higher Harmonics}
\label{sec:observing}
When a gravitational wave signal from a binary merger is observed, it is natural to ask whether the higher multipoles have been observed.  Typically, the searches that identify events do not use higher harmonics to extract events from the data  \citep{Usman:2015kfa, Nitz:2017svb, Hanna:2019ezx} (although see \cite{Harry:2017weg} for ways to incorporate them).  However, the parameter estimation routines do incorporate the higher harmonics into the recovery of parameters, and a natural way to ask whether higher harmonics have been observed is to calculate the Bayes factor (or odds ratio) between parameter recovery with and without higher multipoles in the waveform \citep{LIGOScientific:2020stg, Abbott:2020khf}.  

\subsection{Measured SNR in higher modes}

Here, we consider the SNR contained in the higher modes.  As in \eqn{eq:two_harmonics}, we consider only two harmonics, the 22 harmonic and a single additional harmonic.  Since the 22 harmonic has been identified, it is straightforward to calculate the SNR in the $\ell m$ harmonic, by generating the $\tilde{h}_{\ell m}$ waveform, with the same masses, spins and arrival time, and filtering it against the data.  If the overlap between the $\ell m$ and 22 harmonics is non-zero, then this will pick up power contained in the 22, and it is necessary to remove it by first computing the orthogonal component, 
\begin{equation}
	\tilde{h}^{\perp}_{\ell m} = \tilde{h}_{\ell m} - \frac{1}{|\tilde{h}_{22}|^{2}} \left[ 
	( \tilde{h}_{\ell m} | \tilde{h}_{22}) \tilde{h}_{22} + ( \tilde{h}_{\ell m} | i \tilde{h}_{22}) i \tilde{h}_{22}
	\right] \, .
\end{equation}
Filtering that against the data, $s$, gives
\begin{equation}\label{eq:hm_matchedfilter}
	(\rho_{\ell m}^\perp)^{2} = \frac{1}{ |\tilde{h}^{\perp}_{\ell m}|^{2}} \left[ 
	(s | \tilde{h}^{\perp}_{\ell m})^2  + (s | i \tilde{h}^{\perp}_{\ell m})^2 \right] \, .
\end{equation}
When the parameters of the waveform are known, or have been inferred through parameter estimation, we can calculate the expected SNR in the $\ell m$ mode as
\begin{align}
    \hat{\rho}_{\ell m}^\perp = \hat{\rho}_{\ell m} \sqrt{1 -  O(\ell m, 22)^2} \, .
\end{align}

\begin{figure}[t] 
\includegraphics[width=0.99\linewidth]{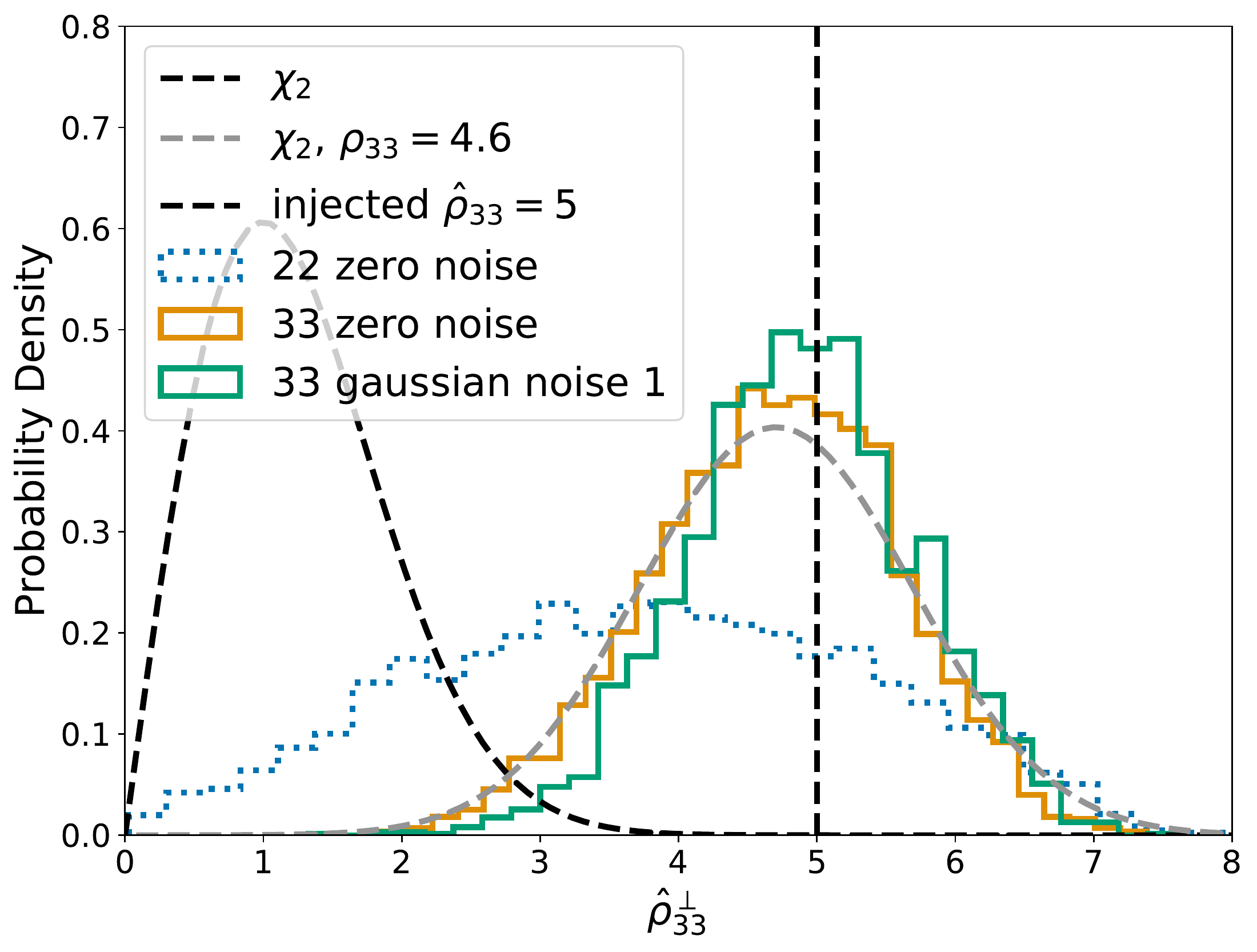} 
\caption{Posterior probability distribution for the orthogonal optimal signal-to-noise ratio of the 33 multipole. The simulated waveform corresponds to system with $m_1=40 M_{\odot}$, $m_2=10 M_{\odot}$ and $\cos\iota=0.7$. The dotted histogram shows the posterior recovered with a waveform model including only the dominant 22 harmonic. The solid histograms show the posterior when the 33 multipole is also included in the model. The dashed curves show $\chi_2$ distributions. We see good agreement between the inferred matched filter SNR and the injected SNR. 
}
\label{fig:rho_33_dist}
\end{figure}

\begin{figure*}[t]
    \includegraphics[width=0.99\linewidth]{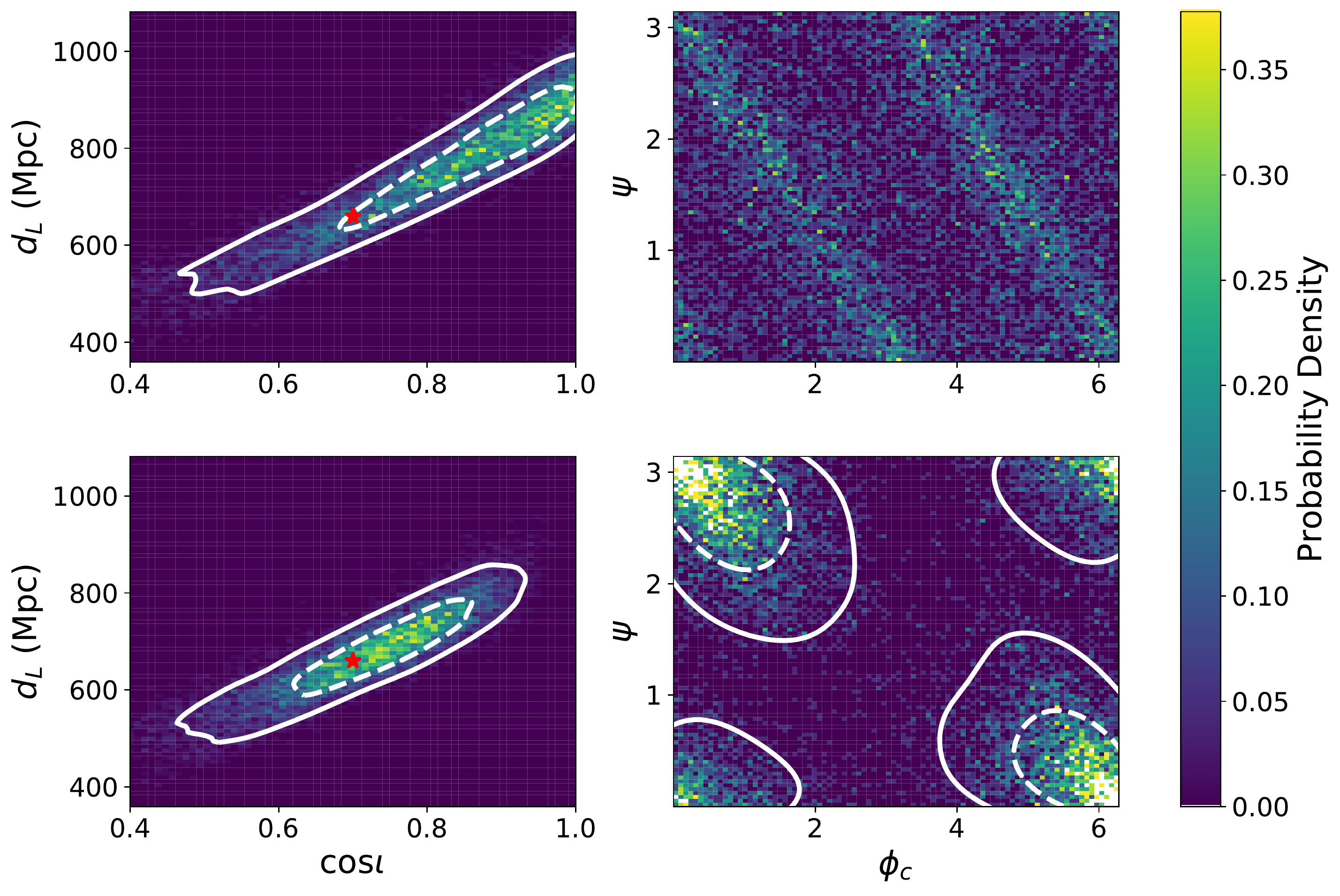}
    \caption{2D Posterior probability distribution for \textit{left:} inclination and distance, \textit{right:} polarization and phase at coalescence for a signal model containing \textit{top:} only the dominant 22 multipole and \textit{bottom:} the 22 and 33 multipoles. The simulated waveform corresponds to a system with $m_1=40 M_{\odot}$, $m_2=10 M_{\odot}$ and $\cos\iota=0.7$. The solid (dashed) white contours denote 90\% (50\%) credible regions.  These are not shown for polarization--phase for the 22 waveform, due to the clear degeneracy.
    }
    \label{fig:degeneracies}
\end{figure*}

In \fig{fig:rho_33_dist} we show the inferred posterior probability distribution for $\hat{\rho}_{33}^\perp$ for a binary with masses $m_1=40 M_\odot$, $m_2=10 M_\odot$ inclined at $\cos\iota=0.7$ ($\iota \approx 45^{\circ}$) and with $\rho_{22} = 20$ under a variety of assumptions for signal and model.
\footnote{
	All parameter estimates reported in this paper were obtained with LALInference \cite{Veitch:2014wba} assuming a HLV network with the sensitivities achieved during O3 \citep{PSDs}. 
}
 For the simulated signal the relative amplitude of the 33 mode is $\alpha_{33} \approx 0.2$, and the orientation factor is $R_{33} = 1.4$, which implies $\hat{\rho}_{33} \approx 5$. We see the recovered distribution of $\hat{\rho}_{33}^\perp \sim 5$ matches well with the simulated value -- peaking at 5.  In addition, we can also \textit{infer} the power in the 33 harmonic even when we use only the dominant 22 harmonic to recover the parameters of the waveform.  Unsurprisingly, the distribution of $\hat{\rho}_{33}^\perp$ no longer matches well with the simulated value and now spreads over a broad range from 0 to 8.  In this case, it seems clear that the 33 mode has been observed, as its inclusion leads to a significant change in the inferred SNR in the 33 mode.  

In \fig{fig:degeneracies} we show the inferred posterior probability distributions of inclination, distance, polarization and phase at coalescence using waveform models that do/do not include the higher harmonics.  Although the binary is generated with the orbital plane inclined at an angle of $\iota \approx 45^{\circ}$, using only the 22 mode, the system is recovered consistent with face-on, due to well-known degeneracies between distance in inclination \citep{Usman:2018imj}.  Consequently, the only well measured quantities are the amplitude and phase of the circularly polarized waveform that is recovered: $A_{22}\approx\frac{\cos\iota}{D_L}$ and $\phi_{22}\approx\psi + \phi_c$, with the inclination bounded to by $0 \le \iota \le 45^{\circ}$.   When the 33 harmonic is added, the degeneracy is broken and the distance, inclination, polarization and phase are all measured with good accuracy.


\subsection{Expectation due to noise}

The question, then, is whether an observed SNR in a given higher mode is evidence that the higher mode has been observed, or if this is to be expected due to noise alone.  In order to answer this, we calculate the expected distribution of $\hat{\rho}_{\ell m}^\perp$ under some simplifying assumptions. Specifically, we consider the scenario where the 22 measurement has already fixed the parameters which determine the phase evolution of the binary (primarily the chirp mass, but also a combination of aligned spin and mass ratio \citep{Baird:2012cu}), the time of arrival and sky location of the system.  Furthermore, we assume that the $\ell m$ mode is the second most significant (in many cases, this is the 33 mode), and other modes do not contribute significantly.
    
As shown in \fig{fig:degeneracies}, there are degeneracies in the measurement of distance/inclination and polarization/phase when observing only the 22 harmonic.  The amplitude of the higher modes, and in particular the 33 and 44 modes, varies significantly over the range $0^{\circ} \le \iota \le 45^{\circ}$ and can therefore be treated as unconstrained.  Similarly, the overall phase of these modes differs from the 22 by a factor of $(m - 2) \phi_{c}$ and is therefore unconstrained by the measurement of the 22 harmonic.  
Another way to see this is to look at the posterior probability distribution for the 33 amplitude inferred when using a 22-only waveform model (see dotted histogram in \fig{fig:rho_33_dist}). The distribution is broad and has support across a large range of $\hat{\rho}_{33}^\perp$. 
We note that this argument will only hold for the subdominant harmonic: once the amplitude of a second harmonic is fixed, the four orientation parameters of the binary are determined and, consequently, the amplitude of the remaining harmonics is significantly constrained.

We are interested in obtaining the expected distribution for $\hat{\rho}_{\ell m}$\footnote{For simplicity of presentation, we drop the $^{\perp}$ from the equations in the remainder of the section.  Where the harmonic has overlap with the 22 mode, the calculation should be understood to be performed with the orthogonal component.} under the noise-only hypothesis (i.e. only power in the 22 mode).  In this case, we are fitting the data with a template waveform
\begin{equation}
h = a \tilde{h}_{\ell m} + i b \tilde{h}_{\ell m} +  c \tilde{h}_{22} + i d \tilde{h}_{22} \, .
\end{equation}

%
Where $h_{\ell m}$ and $i h_{\ell m}$ are the two phases of the waveform of the $\ell m$ harmonic, a and b control the overall amplitude of this harmonic, and $\tilde{h}_{22}$, $c$ and $d$ give the contribution of the dominant harmonic to the waveform. We are interested in the expected distribution of $a$ and $b$ when there is no power in higher modes.  Based upon the discussion above, we choose a uniform prior $\pi(a, b)$ on $a$ and $b$.   In what follows we neglect the terms related to the dominant harmonic as they are unaffected, to the level of our approximation, by the presence of the higher modes.
The posterior will be
\begin{equation}
p(a, b|s) \propto \Lambda(a, b) \pi(a,b) \, .
\end{equation}
where the likelihood of a signal s given the amplitudes a, b and Gaussian noise is
\begin{align}
\label{eq:like}
\Lambda(s | a, b) &\propto \exp\left[ - \frac{1}{2} \left(s - h(a,b) | s-h(a,b) \right)\right] \, .
\end{align}
Using polar variables $\hat{\rho}_{\ell m}=\sqrt{a^2 + b^2}$ and $\hat{\phi}_{\ell m}=\arctan(b/a)$, and assuming a uniform prior we can write the posterior probability distribution for the amplitudes a and b given a signal s as
%
\begin{align*}
p(a, b|s) da db &\propto \Lambda(a, b) da db \\ 
&\propto e^{\left[ a(s|h_{\ell m}) + b (s | i h_{\ell m}) - \frac{a^2 + b^2}{2} \right] }da db \\
&= \hat{\rho}_{\ell m} d\hat{\rho}_{\ell m} d\hat{\phi}_{\ell m} \times \\
& \quad e^{\left[- \frac{ \hat{\rho}_{\ell m}^2 }{2}  + \hat{\rho}_{\ell m} [\cos\hat{\phi}_{\ell m} (s|h_{\ell m}) +  \sin \hat{\phi}_{\ell m}  (s | i h_{\ell m}) ]  \right] }
\end{align*}
%
Defining the matched filter signal-to-noise ratio, $\rho_{\ell m}$ as in Eq. (\ref{eq:hm_matchedfilter}) and the phase
%
\begin{equation}
\phi_{\ell m}=\arctan\frac{(s | ih_{\ell m})}{(s|h_{\ell m})}
\end{equation}
and marginalizing over $\hat{\phi}_{\ell m}$, we obtain 
%
\begin{align}
p(\hat{\rho}_{\ell m}|s)  & \propto \hat{\rho}_{\ell m} e^{\left[- \frac{ \hat{\rho}_{\ell m}^2}{2}  \right] } \int_0^{2 \pi} 
e^{\left[ \hat{\rho}_{\ell m} \rho_{\ell m} \cos(\hat{\phi}_{\ell m} - \phi_{\ell m}) \right]} d\hat{\phi}_{\ell m} 
\nonumber \\
&\propto  \hat{\rho}_{\ell m} \exp\left[- \frac{\hat{\rho}_{\ell m}^2 + \rho_{\ell m}^2 }{2}  \right]  I_0(\hat{\rho}_{\ell m} \rho_{\ell m} )
\label{eq:nc_chi}
\end{align}
%
Where $I_0$ is the modified Bessel function of the first kind, and we have used the fact that $(s|s)=\rho_{\ell m}^2  +$ const. in the approximation of a fixed the phase evolution.%
\footnote{
    The data, $s$, is composed of components parallel, and perpendicular to the two filters $h_{\ell m}$ and $i h_{\ell m}$,
    while $\rho_{\ell m}^2$ picks out the parallel components.
}
We recognize \eqn{eq:nc_chi} as the non-central chi distribution with 2 degrees of freedom and non-centrality parameter equal to $\rho_{\ell m}$. In the absence of power in the higher harmonics, the probability distributions for the filters $(s|h_{\ell m})$ and $(s | i h_{\ell m})$ are zero-mean, unit-variance gaussians and $\rho_{\ell m}$ is chi-distributed with 2 degrees of freedom. 

\begin{figure}[t]
\includegraphics[width=0.99\linewidth]{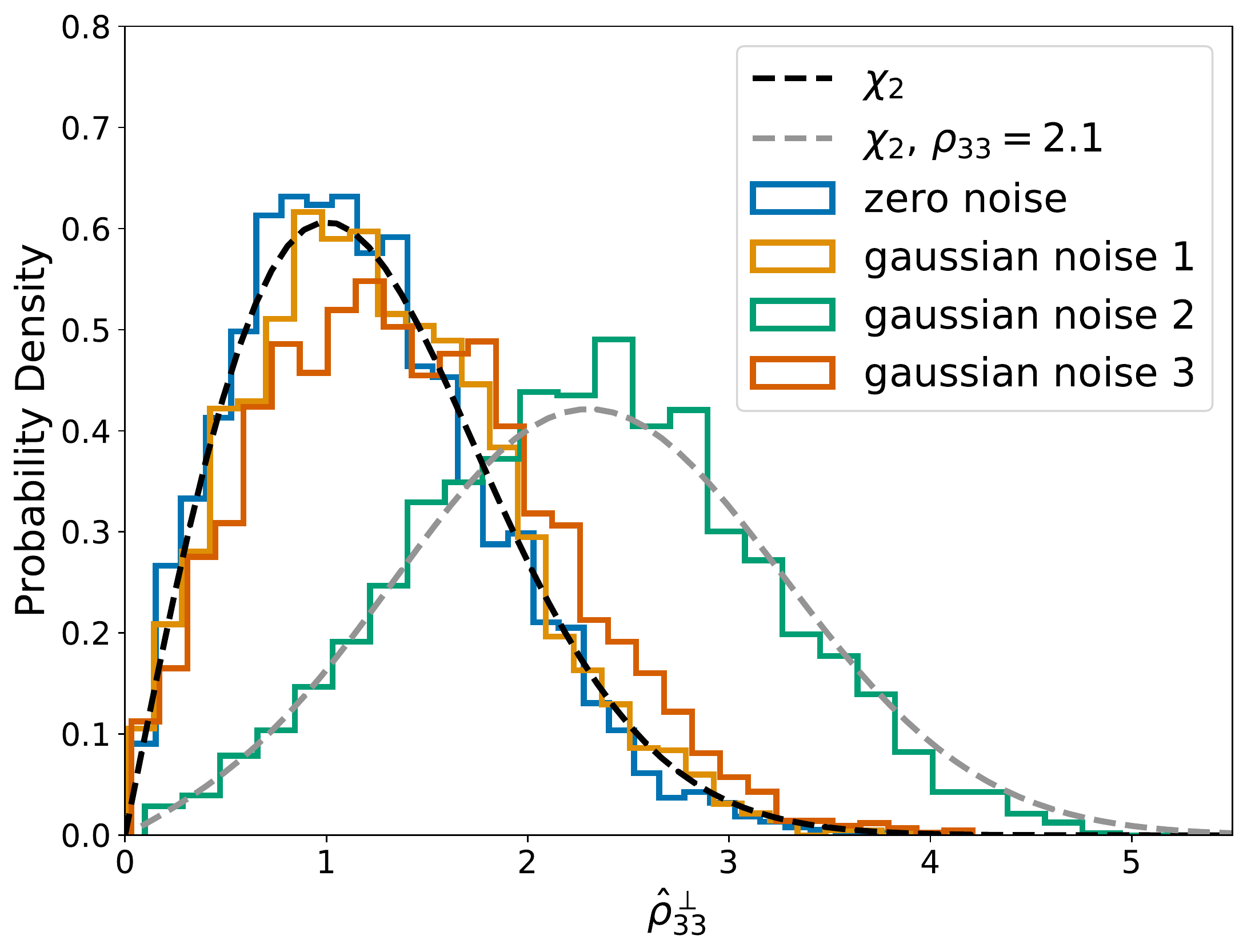}
\caption{Posterior probability distribution for the orthogonal optimal signal-to-noise ratio of the 33 multipole, when the injected signal only contains the 33 multipole for a variety of models and noise realizations. The injected parameters are $m_1=40 M_\odot$, $m_2=10 M_\odot$ at $\cos\iota=0.7$.
}
\label{fig:rho_33_noise}
\end{figure}

\subsection{Observation of higher modes}

In \fig{fig:rho_33_noise} we show the distribution of $\hat{\rho}_{33}$, in the absence of a signal in the 33 mode.  First, we have the recovered distribution when performing parameter estimation on a signal containing only the 22 harmonic and the zero instance of the noise distribution.  Based upon the calculation above, we expect this to follow the $\chi$ distribution with two degrees of freedom, and we see that it does.  We also show the distribution for three different instances of Gaussian noise.  In each of these cases, the distribution is expected to follow a non-central $\chi$ distribution, where the non-centrality parameter is given by the matched filter SNR in the 33 mode -- in this case, there is no signal and any power is simply due to noise.  For two of the noise realizations, there was minimal power in the 33 harmonic and the $\hat{\rho}_{33}$ distribution matches closely with the zero-noise case.  However, in the third noise realization, the SNR in the 33 mode is higher, and the mode of the distribution is moved significantly away from zero.  Finally, we return to \fig{fig:rho_33_dist} and note that the distribution of $\hat{\rho}_{33}$ for the signal containing higher modes matches well with expectation -- a non-central $\chi$ distribution with non centrality parameter $4.6$.

Using these results, we propose a straightforward test of whether the higher harmonics have been observed.  We have argued that the matched filter SNR in the \textit{second most significant} harmonic, in the absence of signal, will be well approximated by a $\chi$ distribution with two degrees of freedom.  We expect Gaussian noise to produce an SNR greater than 2.1 less than 10\% of the time and therefore suggest a simple criterion to be that $\rho_{\ell m} > 2.1$ signifies the observation of power in the higher modes.  
The estimate of $\rho_{\ell m}$ can be obtained either by matched filtering, or by fitting the measured distribution of $\hat{\rho}_{\ell m}$ based upon parameter estimation and obtaining the non-centrality parameter.  Based on this criteria, our third noise trial would show marginal evidence for presence of the 33 harmonic.  This prescription can be easily extended to a criterion for confident detection of the higher harmonics: a ``5-sigma'' observation could correspond to $\rho_{\ell m} > 5.3$.  In \fig{fig:alpha_lm}, we have added contours at values of $\alpha_{\ell m} = 2.1/20$ and $5.3/20$.  These indicate the approximate boundaries in the mass space where higher harmonics would be marginally/confidently observed for a signal at SNR = 20.  Of course, the actual higher mode SNR will depend also on the orientation factor $R_{\ell m}$, which varies between 0 and 2, with a median value around $1$ for the 33 and 44 modes.

An alternative method of establishing the observability of higher harmonics is to compare the Bayes factor (or evidence) between the a waveform model additionally containing the 33 mode and a model with only the 22 multipole.  The difference in Bayes factor, obtained by marginalizing the likelihood \cite{Veitch:2014wba}, between the two parameter estimation runs (with and without 33 mode) is $\log_{10} \mathrm{BF} = 4.5$.  The SNR in the higher harmonics leads to an increase of the likelihood by a factor of $\approx e^{\rho^{2}/2}$ and, as an initial approximation, to a log Bayes factor of $4.6$ ($\log_{10}$ of the increase in the likelihood).  For a more accurate comparison, we should also account for the prior distribution, as well as the width of the posteriors.  Since both the 22 only and higher harmonic waveforms are described by the same parameters, the priors are unchanged.  However, as is clear from \fig{fig:degeneracies}, the posterior is significantly more peaked when the higher harmonics are included.  The improved constraints from the 33 mode reduce the prior volume by a factor of $\sim 2$ in the distance inclination plane (assuming a uniform in volume prior), and a factor of $\sim 5$ in the polarization phase plane. This implies the Bayes Factor based purely on the increased likelihood be reduced by a factor of $\sim 10$, equivalent to reducing the log Bayes Factor by one to $3.6$.  This is in reasonable, but not perfect agreement with the full parameter estimation result.

\section{Higher Harmonics in a Population of Binary Mergers}
\label{sec:population}
\begin{figure*}[t] 
    \includegraphics[width=0.32\textwidth]{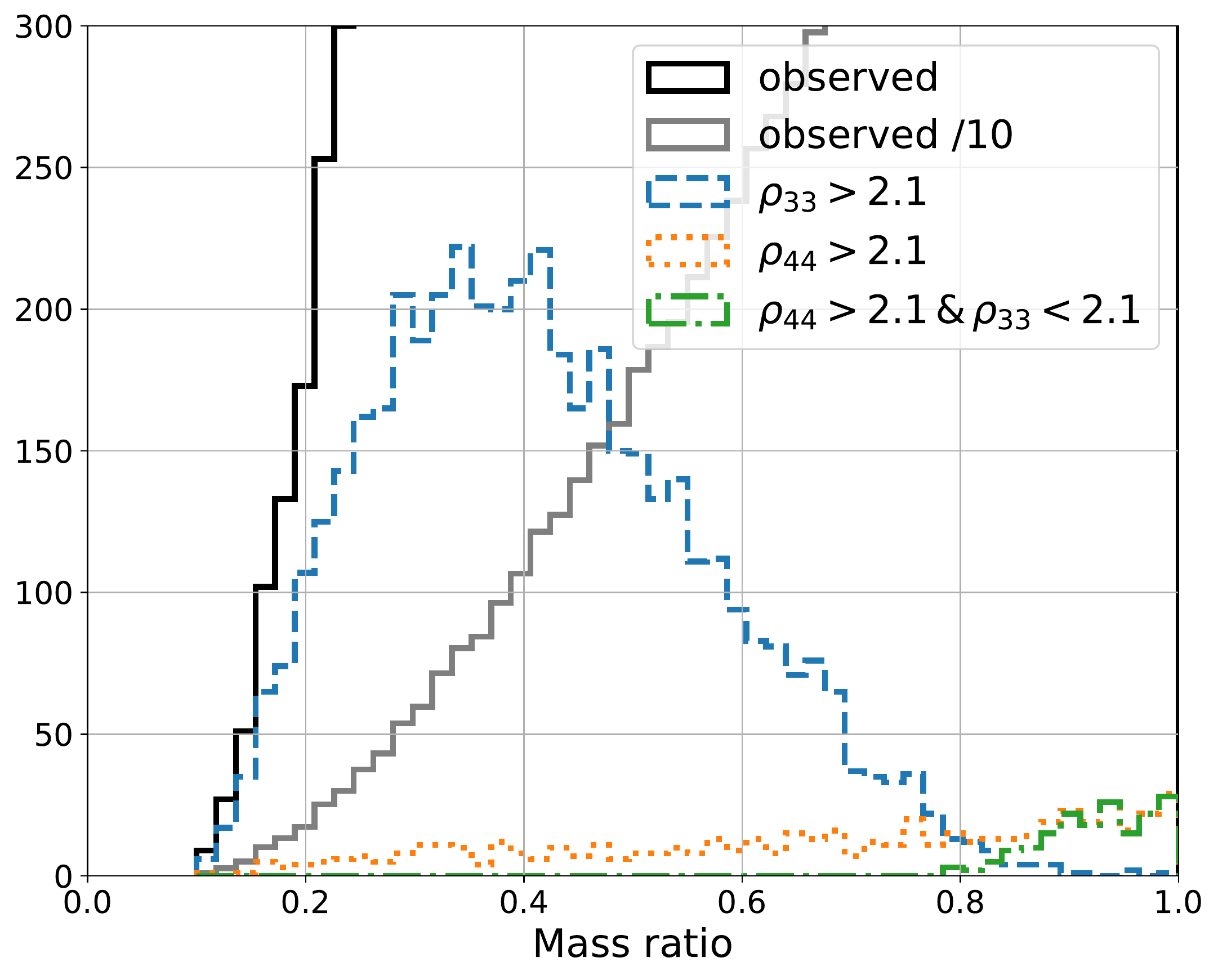} 
    \includegraphics[width=0.32\textwidth]{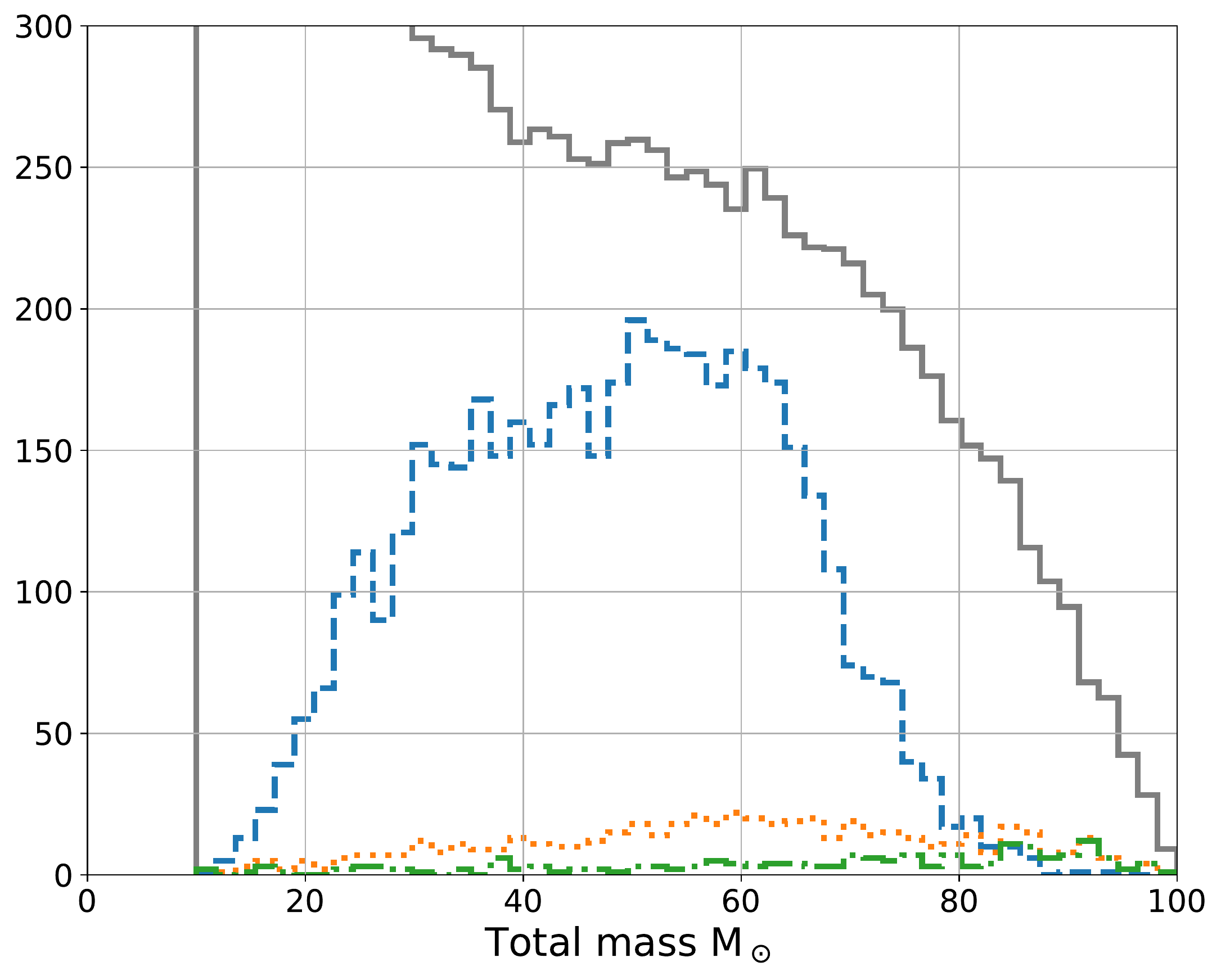} 
    \includegraphics[width=0.32\textwidth]{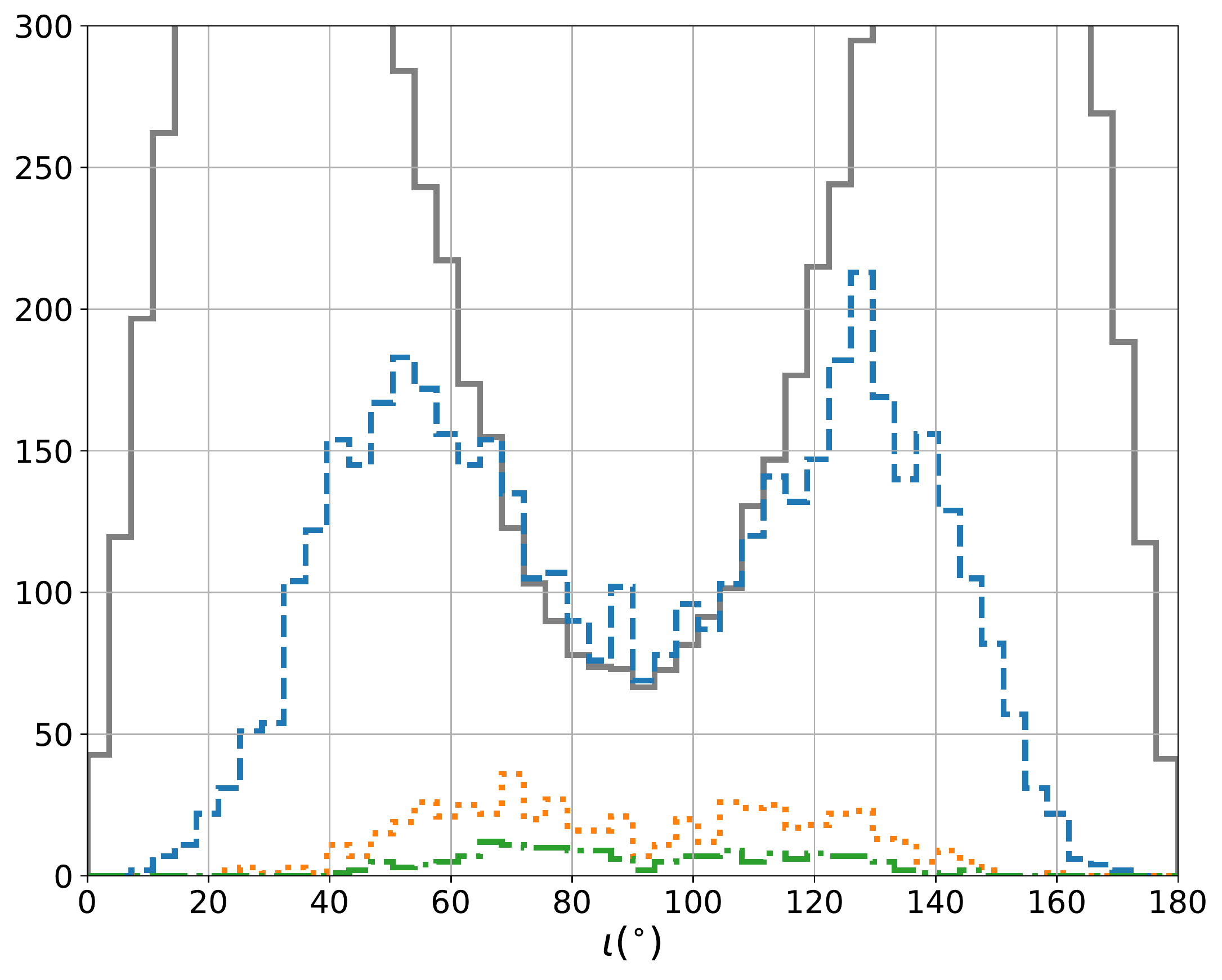} 
    \caption{The mass and orientation distribution for a population of black hole binaries, and for that subset of systems for which the 33 or 44 harmonics are visible. The population is modelled as a power law in masses, with isotropic distribution of orientations, further details of the population are given in the text.  The distribution is shown as a function of \textit{Left:} mass ratio; \textit{Middle:} total mass' \textit{right:} orientation.  The subset of sources for which the 33/44 mode is observable are shown by the dashed/dotted lines respectively, and those with observable 44 mode but not 33 are shown by the dot-dashed line.  We show the observed population divided by 10 as a grey solid line, and on the mass ratio plot, without reweighting as a black line.      }
    \label{fig:population}
    \vspace{10pt} 
\end{figure*}

Here, we consider the likelihood of observing the higher harmonics in signals drawn from a population.  To do so, we generate a large number of potential signals from a given population and assess which would be observable above a given threshold and, of those, which would have sufficient power in the 33 and/or 44 harmonics for them to be observable (above the threshold of $\rho_{\ell m} = 2.1$).  We choose a mass distribution of black holes in binary systems where the mass of the more massive black hole is taken from a power-law distribution $p(m_{1}) \propto m_{1}^{-\alpha}$ and choose the power law parameter of $\alpha=-2.35$, while restricting the mass to lie in the range [5, 50] M$_\odot$;  the distribution for $m_{2}$ is taken to be uniform in the range $[5 M_{\odot},  m_{1}]$.  The spins of the individual black holes are assumed isotropically distributed, with low spin magnitudes (the magnitude is a triangular distribution peaked at spin magnitude of zero and falling to zero at maximal spin) \citep{Tiwari:2018qch}. 

In \fig{fig:population} we show the subset of this population which would be detectable with the HLV network operating at the sensitivities achieved during O3 \citep{PSDs}. More pertinently, we also plot the subset from this detected population which result in gravitational waves with a measurable signal in the two loudest subdominant multipoles. Overall around 3.5\% of binaries are expected to have sufficient power in the higher harmonics for them to be observed.  Of these, the vast majority will have an observable 33 mode (3.4\%), with the 44 mode observable in 0.4\% of binaries, but for the majority of these, the 33 mode will also be observable.  Only one observable event in 1,000 from this population is expected to have an observable 44 mode \textit{but not} observable 33.  

The higher harmonics are preferentially observable in signals with unequal masses and for sources for binaries which are significantly inclined.  In particular, for binaries with mass ratio between 3:1 and 10:1, the majority of signals will have observable higher modes, and even at a mass ratio of 2:1, around 10\% of binaries will have an observable higher mode.  Convolving the observed distribution with the fraction of binaries with significant higher modes gives a peak of observed HM signals around a mass ratio of 3:1. Interestingly, for binaries close to equal mass, it is the 44 mode which is more likely to be observed, and \textit{essentially all} binaries where the 44 but not 33 is observed have close to equal masses (between 1:1 and 5:4).  


\section{Discussion}
\label{sec:discussion}
We have explored the relative significance of the higher gravitational wave harmonics in binary merger signals.  For simplicity, we have decomposed the harmonics into an overall amplitude --- dependent upon the masses and spins of the system --- and an orientation-dependent term --- dependent upon the inclination and polarization.  This allows us to easily identify the most significant modes, and the regions of parameter space where they are most likely to be observable.  As is well known \citep{Kidder:2007rt, Mishra:2016whh, Kalaghatgi:2019log}, the higher harmonics are most significant when the binary is observed edge on.  As expected, our orientation amplitudes are largest for edge-on systems although, due to selection effects, we observe that the most likely observed configuration is a binary with axes orientated at around $45^{\circ}$ to the line of sight. In addition, we show that for much of the binary parameter space, the 33 mode will be the most significant sub-dominant harmonic, with an amplitude about one quarter of the 22 mode for a mass-ratio 2 binary (over a broad range of masses).  The 44 mode becomes more significant at higher masses and, although the relative amplitude is less than $0.2$ for much of the parameter space, it is still the most significant sub-dominant harmonic for high-mass systems where the two components have comparable masses

For signals which are close to threshold, it is likely that only one additional mode will be clearly observable.  Thus, for simplicity, we have introduced and observability criterion for the second harmonic.  In many cases, the amplitude and phase of the second harmonic is largely unconstrained by the observation of the 22 mode: there are often large degeneracies between the measurement of the distance and inclination of the binary and also the polarization and phase \citep{Usman:2018imj}.  Consequently, in the absence of a signal, the power in the second most significant mode will be $\chi^{2}$ distributed with two degrees of freedom, corresponding to the unknown amplitude and phase of the mode.  If there is power in the higher harmonic, the distribution will be non-central $\chi^{2}$, where the non-centrality is given by the SNR in the higher mode.  We have performed a series of simulations that demonstrate this expectation is valid.  Using this simple observation, we have introduced a criterion for observation of power in a higher mode: if the observed SNR in the mode is above 2.1, then this is unlikely to occur due to noise alone so there is marginal evidence of a higher mode signal, while an SNR $>5.3$ would provide strong (``5-sigma'') evidence.

We have identified regions in the parameter space where higher harmonics are most likely to be observed.  These regions are those where higher harmonics are likely to be observed, but also which are relatively common in the underlying population of observed gravitational wave signals \citep{O2RatesPops}.  We find that these correspond to signals with mass ratios between 2:1 and 5:1 --- for more equal masses, the higher harmonics are too weak, more unequal mass binaries are thought to be rarer.  Furthermore, the most likely orientation is for the axis to be inclined at between $30^{\circ}$ and $60^{\circ}$ to the observer --- less inclined systems have insufficient power in the higher harmonics while more inclined systems have a weaker overall emission.  

There are several applications of the work presented here.  As already mentioned, the criterion for observability of higher harmonics has been in used, along with other methods \cite{Roy:2019phx}, in establishing the presence of power in the 33 mode in the observed signals GW190412 and GW190814 \citep{LIGOScientific:2020stg, Abbott:2020khf}.  Furthermore, the method can be used in a straightforward way to determine whether it is likely that the higher harmonics will be observable in a given system, and we have provided an example in the population study presented in section \ref{sec:population}.  This is directly applicable to signals observed using a search for the dominant harmonic.  Based upon the observed parameters, we can calculate the expected power in the higher modes and identify the expected SNR.  If significant SNR is expected in higher harmonics, then it becomes worthwhile to undertake the (computationally costly)  parameter estimation with a higher-mode waveform.  This will lead either to the observation of  higher modes, and the subsequent improvement of parameter measurement, or the non-observation of higher modes and subsequent restriction of the binary parameters to regions of the parameter space where the higher harmonic amplitudes are low.

While the method introduced here is straightforward, there are several clear limitations.  Most obviously, the discussion has limited attention to a single observable harmonic.  In many cases, this will be a reasonable approximation as there will be one harmonic which is significantly larger than the others (as can be seen from \fig{fig:alpha_lm}).  Furthermore, from simple parameter counting, it seems likely that the observation of a single higher harmonic will be sufficient to significantly improve parameter recovery, most notably the binary orientation.  Nonetheless, the observation of additional modes will provide additional improvements.  For a detailed understanding of the impact of all of the higher harmonics, a full, Bayesian parameter estimation exploration of the issue will be necessary \cite{Kalaghatgi:2019log}. Additionally, throughout the paper, we have used a single waveform model, IMRPhenomHM \citep{London:2017bcn} and checked for consistency with a more recently updated model IMRPhenomXHM \citep{Garcia-Quiros:2020qpx}; but results are likely to vary somewhat with other models of the higher harmonics (for example, \cite{Varma:2018mmi, Cotesta:2020qhw, Nagar:2020pcj}). Finally, we have restricted attention throughout the paper to non-precessing systems.  Recently, \citep{Fairhurst:2019vut, Fairhurst:2019srr}, an analysis similar to the one presented here was performed on precessing systems, again with a focus on the observability of the two dominant harmonics.  For systems where both higher harmonics and precession have an significant impact on the waveform, it will be necessary to combine these approaches to develop a straightforward categorization of precessing systems with observable higher harmonics.

\section*{Acknowledgments}
The authors have benefitted from many valuable discussions with Edward Fauchon-Jones, Cecilio Garc{\'i}a-Quir{\'o}s, Rhys Green, Eleanor Hamilton, Mark Hannam, Charlie Hoy, Sebastian Khan, Lionel London, Frank Ohme, Vaibhav Tiwari and Vivien Raymond. SF and CM acknowledge support from the Science and Technology Facilities Council (STFC) grant ST/L000962/1, European Research Council Consolidator Grant 647839ST/L000962/1. Finally, the authors are grateful for computational resources provided by the Cardiff University and LIGO Laboratory and supported by STFC grant ST/I006285/1 and National Science Foundation Grants PHY-0757058 and PHY-0823459.

\appendix

\section{Spin-weighted spherical harmonic polarizations}
\label{sec:Ylm}
The general form for the spin-weighted spherical harmonics is
\begin{widetext}
\begin{equation}
{}_sY_{l m} (\iota, \phi_c) = \left(-1\right)^m \sqrt{ \frac{(l+m)! (l-m)! (2l+1)} {4\pi (l+s)! (l-s)!} } \sin^{2l} \left( \frac{\iota}{2} \right) \times\sum_{r=0}^{l-s} {l-s \choose r} {l+s \choose r+s-m} \left(-1\right)^{l-r-s} e^{i m \phi_c} \cot^{2r+s-m} \left( \frac{\iota} {2} \right) , \nonumber
\end{equation}
\end{widetext}
which can be written in terms of the wigner d-functions $d^{lm}_{-s}(\iota)$ (implicitly defined here)
\begin{equation}
{}_sY_{l m} (\iota, \phi_c) =  \sqrt{ \frac{ (2l+1)} {4\pi} } d^{lm}_{-s}(\iota) e^{i m \phi_c}  .
\label{eq:ylm_def}
\end{equation}
They have the following symmetries 
\begin{align*}
{}_s\bar Y_{l m} &= \left(-1\right)^{s+m}{}_{-s}Y_{l(-m)}\\
{}_sY_{l m}(\pi-\iota,\phi_c+\pi) &= \left(-1\right)^l {}_{-s}Y_{l m}(\iota,\phi_c) .
\end{align*}
The spin-weighted spherical harmonics for the modes we are interested in are
\begin{align*}
{}_{-2}Y_{22} &= \frac{1}{2} \sqrt{\frac{5}{\pi }} e^{2 i \phi_c }  \cos ^4\left(\frac{\iota }{2}\right) \numberthis \\
{}_{-2}Y_{2-2} &= \frac{1}{2} \sqrt{\frac{5}{\pi }} e^{-2 i \phi_c } \sin ^4\left(\frac{\iota }{2}\right) \\
{}_{-2}Y_{21} &= \frac{1}{2} \sqrt{\frac{5}{\pi }} e^{i \phi_c } \cos ^2\left(\frac{\iota }{2}\right) \sin (\iota ) \\
{}_{-2}Y_{2-1} &= \frac{1}{2} \sqrt{\frac{5}{\pi }} e^{-i \phi_c } \sin ^2\left(\frac{\iota }{2}\right) \sin (\iota ) \\
{}_{-2}Y_{33} &= \frac{1}{2} \sqrt{\frac{21}{2 \pi }} \left(-e^{i 3 \phi_c )}\right) \cos ^4\left(\frac{\iota }{2}\right)  \sin (\iota ) \\
{}_{-2}Y_{3-3} &= \frac{1}{2} \sqrt{\frac{21}{2 \pi }} e^{-i 3 \phi_c } \sin ^4\left(\frac{\iota }{2}\right) \sin (\iota ) \\
{}_{-2}Y_{44} &= \frac{3}{4} \sqrt{\frac{7}{\pi }} e^{4 i  \phi_c } \cos ^4\left(\frac{\iota }{2}\right) \sin ^2(\iota )\\
{}_{-2}Y_{4-4} &= \frac{3}{4} \sqrt{\frac{7}{\pi }} e^{-4 i  \phi_c } \sin ^4\left(\frac{\iota }{2}\right) \sin ^2(\iota ) \, .
\end{align*}
We can write the gravitational wave polarizations as a sum of these spherical harmonics with coefficients $h_{lm}$
\begin{equation}
h_+ - i h_{\times} = \sum_{l\geq2} \sum_{m=-l}^{l} {}_{-2}Y_{lm}(\iota, \phi_c) h_{lm} \, .
\label{eq:decomp}
\end{equation}
%
Three properties of $h_{lm}$ help to simplify \eqn{eq:decomp}. Firstly, specializing to planar (i.e. non-precessing) 
binaries allows us to write $h_{l-m}=(-1)^l h_{lm}^*$\citep{Blanchet:2013haa}. Secondly, in the frequency domain, 
$\tilde{h}^*_{l-m}(f) = \tilde{h}_{lm}(-f)^*$. 
%
%
Finally we make the further approximation \citep{Marsat:2018oam} that if we only care about the waveform in direction $\hat{n}$ we can neglect one side of the frequency spectrum, depending on the sign of m.  This approximation is valid in particular where the stationary phase approximation has been used. We therefore assume, with the sign convention on the Fourier transform as $\tilde{h}(f)=\int dt h(t) e^{+i2\pi ft}$, that
\begin{align}
\tilde{h}_{lm}(f) \simeq 0 \begin{cases}  f>0, m<0 \\  f<0, m>0 . \end{cases} 
\end{align}

With these three properties we can obtain explicit expressions for the orientation dependence of each of the modes for positive frequencies
\begin{equation}
\begin{split}
h_+ = \frac{1}{2} \sum_{l\geq2} \sum_{m=-l}^{l} & \Big[  {}_{-2}Y_{lm}(\iota, \phi_c) h_{lm} + {}_{-2}Y_{lm}^*(\iota, \phi_c) h_{lm}^* \Big] \\
\tilde{h}_+(f) = \frac{1}{2} \sum_{l\geq2} \sum_{m=-l}^{l} & \Big[ {}_{-2}Y_{lm}(\iota, \phi_c) \tilde{h}_{lm}(f) \Big. \\
& \Big.+  {}_{-2}Y_{lm}^*(\iota, \phi_c) \tilde{h}_{lm}(-f)^* \Big] \\
= \frac{1}{2} \sum_{l\geq2} \sum_{m=1}^{l} & \Big[  {}_{-2}Y_{lm}(\iota, \phi_c) \Big. \\
& \Big. +  (-1)^l  {}_{-2}Y_{l-m}^*(\iota, \phi_c) \Big] \tilde{h}_{lm}(f) 
\end{split}
\end{equation}
and similarly we can show
\begin{equation}
\begin{split}
h_{\times} = \frac{i}{2} \sum_{l\geq2} \sum_{m=-l}^{l} & \Big[ {}_{-2}Y_{lm}(\iota, \phi_c) h_{lm} - {}_{-2}Y_{lm}^*(\iota, \phi_c) h_{lm}^* \Big] \\
\tilde{h}_{\times}(f) = \frac{i}{2}  \sum_{l\geq2} \sum_{m=-l}^{l} & \Big[ {}_{-2}Y_{lm}(\iota, \phi_c) \tilde{h}_{lm}(f) \Big. \\
& \Big. -  {}_{-2}Y_{lm}^*(\iota, \phi_c) \tilde{h}_{lm}(-f)^* \Big] \\
= \frac{i}{2} \sum_{l\geq2} \sum_{m=1}^{l} & \Big[ {}_{-2}Y_{lm}(\iota, \phi_c) \Big. \\
& \Big. -  (-1)^l  {}_{-2}Y_{l-m}^*(\iota, \phi_c) \Big] \tilde{h}_{lm}(f) \, .
\label{eq:Ylm_factored}
\end{split}
\end{equation}
where in both cases, we have neglected the $m=0$ terms in the sums as they are not considered in the models we have used.  Finally, we note that we have used a different normalization convention in the main text (\eqn{eq:Alm_polarizations}) than the one typically used in the spin-weighted spherical harmonic decomposition described in this Appendix.  This has no impact on the results, but merely changes the values of $\alpha_{\ell m}$ and $R_{\ell m}$ while maintaining the same values of the SNR in the higher modes.

\section{Derivation of $p(R_{lm})$}
\label{sec:derive_R}
We now derive the probability distributions in \fig{fig:R_dist}. 
Assuming no preferred orientation for binaries in the universe, the probability density function for $\cos\iota$, $p(\cos\iota)$, is
\begin{equation}
    p_{\mathrm{univ}}(\cos\iota) = \tfrac{1}{2}
\end{equation}
However, binaries which emit primarily in the 22 multipole radiate 
most powerfully in the direction perpendicular to the orbital plane, $|\cos\iota|\sim1$.  Consequently, the horizon for the subset of these binaries which are viewed edge-on is much closer and we preferentially observe face-on binaries. It can be shown \citep{Schutz:2011tw} that the radiated power of the dominant multipole as a function of inclination is
\begin{equation}
    \label{eq:F22}
    F(\iota)^{22} = (A^{22}_+)^2 + (A^{22}_\times)^2 \, ,
\end{equation}
where $A^{22}_{+, \times}$ are defined in \eqn{eq:iota_mode_dependence}.  For a detector sensitive to only one polarization of gravitational wave, the observed power will depend upon the polarization.  This will also be the case for a network with different sensitivities to the two polarization, but not for one equally sensitive to both polarizations of the gravitational wave.  It is possible to approximately marginalize over the polarization distribution and obtain a probability distribution for the inclinations of detected binaries \citep{Schutz:2011tw} 
\begin{equation}
    p_{\mathrm{det}}(\cos\iota) \propto F(\iota)^{3/2} = (1 + 6 \cos^2 \iota + \cos^4 \iota)^{3/2} \, .
\end{equation}

Using these results, it is straightforward to obtain expressions for the distributions for the expected power in the 33 and 44 multipoles, both for a uniform population of binaries and for those which are observable above a fixed threshold.  The distribution for other multipoles can also be obtained but, since in general $R_{\ell m}$ will depend upon polarization angle, the results will be dependent upon the details of the network and its sensitivity to the two gravitational wave polarizations.  For the 33 and 44 modes, the relative amplitude depends only on the inclination angle $\iota$.


To obtain an expression for the probability distribution for $R_{\ell m}$, we change variables
\begin{equation}
    p(R_{\ell m}) =  \left(\frac{d\cos\iota}{dR_{\ell m}} \right) p(\cos\iota) 
\end{equation}
so that, recalling the functional form of $R_{33}$ and $R_{44}$ from \eqn{eq:Rll}, we obtain

\begin{align}
  p_{\mathrm{univ}}(R_{33}) &= \frac{R_{33}}{4 \sqrt{1 - \left(\tfrac{R_{33}}{2}\right)^{2}} }\nonumber \\
  p_{\mathrm{univ}}(R_{44}) &= \frac{1}{4 \sqrt{1 - \left(\tfrac{R_{44}}{2}\right)} } \, .
\end{align}
Assuming binaries are detected with 22-only waveforms, we can apply the same weighting factor as above in obtaining the distributions for the observed binaries, to obtain
\begin{equation}
    p_{\mathrm{det}}(R_{\ell m}) = \left(\frac{d\cos\iota}{dR_{\ell m}}\right) p_{\mathrm{det}}(\cos \iota)
\end{equation}
which gives
\begin{align}
  p_{\mathrm{det}}(R_{33}) &\propto \left( 8 - 2 R_{33}^{2} + \tfrac{R_{33}^4}{16}\right) p_{\mathrm{univ}}(R_{33})
  \nonumber \\
  p_{\mathrm{det}}(R_{44}) &\propto \left( 8 - 4 R_{44} + \tfrac{R_{44}^2}{4}\right) p_{\mathrm{univ}}(R_{44})
\end{align}
These distributions are plotted in \fig{fig:R_dist}, and discussed in the surrounding text.

%
%
%



\bibliography{gw1204-refs}
\end{document}